\documentclass[pra,bibnotes,twocolumn]{revtex4}
\usepackage{amsmath}
\usepackage{graphicx}
\usepackage{color}
\usepackage{hyperref}
\usepackage{graphicx}
\usepackage{amsmath,amssymb}
\usepackage{float}
\usepackage{bm}

\begin{document}
\draft
\def\ds{\displaystyle}
\title{ Non-Hermitian edge burst without skin localizations } 
\author{C. Yuce, H. Ramezani}
\address{Department of Physics, Eskisehir Technical University,  Eskisehir, Turkey }
\address{Department of Physics and Astronomy, University of Texas Rio Grande Valley, Edinburg, Texas 78539, USA}

\date{\today}
\begin{abstract}
In a class of non-Hermitian quantum walk in lossy lattices with open boundary conditions, an unexpected peak in the distribution of the decay probabilities appears at the edge, dubbed edge burst. It is proposed that the edge burst is originated jointly from the non-Hermitian skin effect (NHSE) and the imaginary gaplessness of the spectrum [Wen-Tan Xue et al., Phys. Rev. Lett. 128, 120401 (2022)]. Using a particular one-dimensional lossy lattice with a nonuniform loss rate, we show that the edge burst can occur even in the absence of skin localization. Furthermore, we discuss that the edge burst may not appear if the spectrum satisfies the imaginary gaplesness condition. Aside from its fundamental importance, by removing the restrictions on observing the edge burst effect, our results open the door to broader design space for future applications of the edge burst effect.

\end{abstract}
\maketitle

 \section{Introduction}

The past two decades have witnessed a wealth of promising work in extending the quantum theory to the non-Hermitian domain. Among the many fascinating aspects of non-Hermitian Hamiltonians, the NHSE has recently attracted a great deal of attention \cite{nonbl2,nonbl3}. The NHSE implies that the complex spectrum of a non-Hermitian lattice can be highly sensitive to the boundary conditions, and the eigenstates that are not at the Bloch points (at which the eigenvalues are the same under the periodic and open boundary conditions) are localized at the edge of the open lattice \cite{cy0,10,cy2,ek1,cy3,cy4,cy5,ek4,cy6,cy7,cy8,ek2,ek3,ek5,18c,5,7,13,11,17,18,8a,8acem,16,6b,cypla1,cemyuce2pi,7a,8}. The non-Bloch band theory has been formulated to explain the intriguing features of the NHSE \cite{nbloch}. The NHSE and topology are interconnected \cite{pg09c,pg09,pg09a,ekle1,pg09b,pg09d}, and the NHSE can thus be predicted using the spectral winding number \cite{pg09c}. \\
Quantum dynamics in non-Hermitian systems are believed to be quite different from the standard Hermitian systems. Quantum walk originated from a generalization of the classical random walk has also been extended to non-Hermitian systems \cite{hgt2}. A quantum walker will completely leak out eventually from a bipartite lossy lattice with uniform loss rates  \cite{in1}. The quantum walker in this system is expected to escape predominantly from nearby sites of a starting point that is far from the edges. However, numerical computations show that the decay probability distribution is left-right asymmetric, and a relatively large peak in the loss probability at the farthest edge from the staring point occurs. More unexpectedly, the relative height of this peak grows with the distance between the starting point and the edge. Originally, it was attributed to topological edge states \cite{in1}, which is questioned in a recent paper \cite{in2}. The appearance of an edge peak (so-called the edge burst) was demonstrated to stem entirely from the interplay between two prominent non-Hermitian phenomena, the NHSE and
imaginary gap closing \cite{in2}. The left-right asymmetry is attributed to the NHSE since all eigenstates are localized at one edge of the system, and the large peak at the edge is due to the imaginary gap closing. \\
In this paper, we show that edge burst can occur even if an extensive number of eigenstates are not localized at one edge of the system due to NHSE. We consider the same lattice as \cite{in1,in2} but with non-uniform loss rates. The NHSE disappears due to the nonuniform nature of the loss rates, but imaginary gap closing condition on the spectrum is satisfied. The left-right asymmetry of the decay probability occurs in the system due to the phase difference of the couplings in each unit cell. We also demonstrate that there exist systems with
left-right asymmetric decay probability and without edge burst even if the imaginary gap closing condition on the spectrum is satisfied. 

\section{Quantum Walk}
 
We consider a quantum walker in a tight-binding one-dimensional non-Hermitian lattice with $\ds{N}$ unit cells. The lattice as shown in Fig. 1 is composed of two sublattices $\mathcal{A}$ and $\mathcal{B}$. The non-Hermiticity comes from lossy $\mathcal{B}$ sublattice with nonuniform loss rates. The dynamics of the quantum walker in this lattice obeys the following coupled equations
\begin{eqnarray}\label{anadenk} 
i\frac{d\psi^A_n}{dt} =  t_1\psi^B_n+i\frac{t_2}{2}  (\psi^A_{n-1}-\psi^A_{n+1})  +  \frac{t_2}{2}  (\psi^B_{n-1}  +  \psi^B_{n+1})  \nonumber\\
i\frac{d\psi^B_n}{dt} =  t_1\psi^A_n -i\frac{t_2}{2}  (\psi^B_{n-1}-\psi^B_{n+1})  +  \frac{t_2}{2}  (\psi^A_{n-1} +  \psi^A_{n+1})    \nonumber\\ -i~\gamma_n~  \psi_n^B~~
\end{eqnarray} 
where $\ds{n=1,2,...,N}$, $\ds{\psi_n^{A}(t)}$ and  $\ds{\psi_n^{B}(t)}$ are time-dependent complex field amplitudes in the $\mathcal{A}$ and $\mathcal{B}$ sublattices, respectively, $\ds{t_{1}}$ and $\ds{t_{2}}$ are real positive parameters describing couplings and $\ds{\gamma_n > 0}$ are site-dependent loss rates.\\
Suppose that the quantum walker is initially placed in the $\mathcal{A}$ sublattice at the starting unit cell $\ds{S}$ that is supposed to be close to the right edge. Therefore, the initial conditions are given by $\ds{  \psi_n^A (t=0) =\delta_{n,S}}$ and $\ds{  \psi_n^B (t=0) =0   }$. To study the dynamics, we numerically solve Eq. (\ref{anadenk}) subject to the open boundary conditions and the above initial conditions. During the quantum walk, the walker moves in discrete steps in both sublattices and escape only from lossy $\mathcal{B}$ sites. As $\ds{t\rightarrow\infty}$, the walker completely leaks out from the system. The decay probability that the quantum walker escapes from the leaky $\mathcal{B}$ sublattice with the site number $\ds{n}$ is given by \cite{in1}
\begin{equation}\label{losslinrate} 
P_n= 2~\gamma_n \int_0^{\infty} |  \psi_n^B(t) |^2 ~dt
\end{equation} 
with total decay probability conservation $\ds{\sum_{n=1}^NP_n=1}$.\\
As a special case, the system exhibits left-right asymmetry in decay probabilities when the loss rate is uniform ($\ds{\gamma_n=\gamma}$)\cite{in1,in2}. In this case, $\ds{P_n}$ is maximum at $n=S$ and decreases algebraically in the bulk as $n$ decreases from $S$, and then makes a sharp peak at the left edge (edge burst). On the other hand, $\ds{P_n}$ is very small for $\ds{ n>S  }$. We are interested in exploring the edge burst for the system with nonuniform loss rates. We think that we can enhance or suppress the edge burst by a proper choice of $\ds{\gamma_n}$. For example, one intuitively expects that it can be suppressed if $\ds{ \gamma_n  }$ are large around the starting point $S$ but small around the left edge. Let us specifically suppose that the loss rate increases linearly from the left edge with a constant rate of change $\ds{\gamma}$
\begin{figure}[t]
\includegraphics[width=4.5cm]{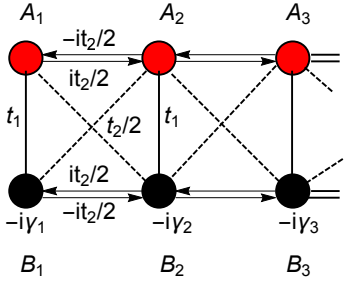}
\caption{The lossy finite tight-binding lattice with two sublattices. Losses occur only in $\mathcal{B}$ subtalltice with nonuniform loss rates $\ds{\gamma_n=\{  \gamma_1,\gamma_2,...,\gamma_N  \}  }$ with $\ds{ N  }$ being the total number of unit cells. A quantum walker starts in the $\mathcal{A}$ sublattice $\ds{A_S}$ that is supposed to be close to the right edge. As $\ds{t\rightarrow\infty}$, the walker completely leaks out from the system.}
\end{figure}
\begin{equation}\label{iyrl0dsf} 
\gamma_n=   {\gamma} ~n  
\end{equation} 
from which one may naively say that the edge burst can be suppressed by thinking that the quantum walker escapes from the system before reaching the left edge of the lattice ($\ds{N>>1}$ and $\ds{S>>1}$). But we numerically see that this is not the case. In fact, the system has radically different behavior than its uniformly lossy analogue and edge burst can be enhanced instead. \\
To quantify the edge burst, we use the relative
height, defined as $\ds{P_1/P_{min}}  $, where $\ds{P_{min}  = \min \{P_1,P_2, ... , P_S
\} }$ is the minimum of $P_n$ between the left edge and the starting point $\ds{S}$. We note that $\ds{P_1/P_{min}   >>  1}  $
and $\ds{P_1/P_{min}   \sim  1  }  $ are the evidence of the existence and absence of the edge burst, respectively \cite{in2}. We can also define another ratio $\ds{P_1/P_{S}  }  $ to compare the decay probabilities at the left edge and starting point. For the uniform loss rate, this ratio is always smaller than $1$, indicating that the quantum walker leaks out more from the starting unit cell than the one at the left edge (Fig.2 (a)). However, the ratio $\ds{P_1/P_{S}  }  $ can be bigger than $\ds{1}$ for the nonuniform loss rate (\ref {iyrl0dsf}) as can be seen from Fig.2 (b-d). This is counterintuitive as it is natural to expect the quantum walker to decay mostly from the starting unit cell, and not from the farthest unit cell, which has also the least loss rate. It seems that the quantum walker reaches the left edge with less losses if we increase $\ds{\gamma}$ and waits there until it decays completely from there. We find that $\ds{P_1}$ grows with $\ds{ \gamma }$, whereas $\ds{P_S}$ decreases with it. At quite large values of the loss rate $(\gamma>3)$, the peak at the starting point disappears. We can also compare the behavior of $\ds{P_n}$ in the bulk with $\ds{  S>n>1 }$. $\ds{P_n}$ decreases algebraically in the bulk as $n$ decreases from $S$ when the loss rate is uniform, $\gamma_n=\gamma$ (Fig.2 (a)). However, it stays almost constant at many sites in the bulk when the loss rate is nonuniform (\ref{iyrl0dsf}) (Fig.2 (b-d)).\\
\begin{figure}[t]
\includegraphics[width=4.25cm]{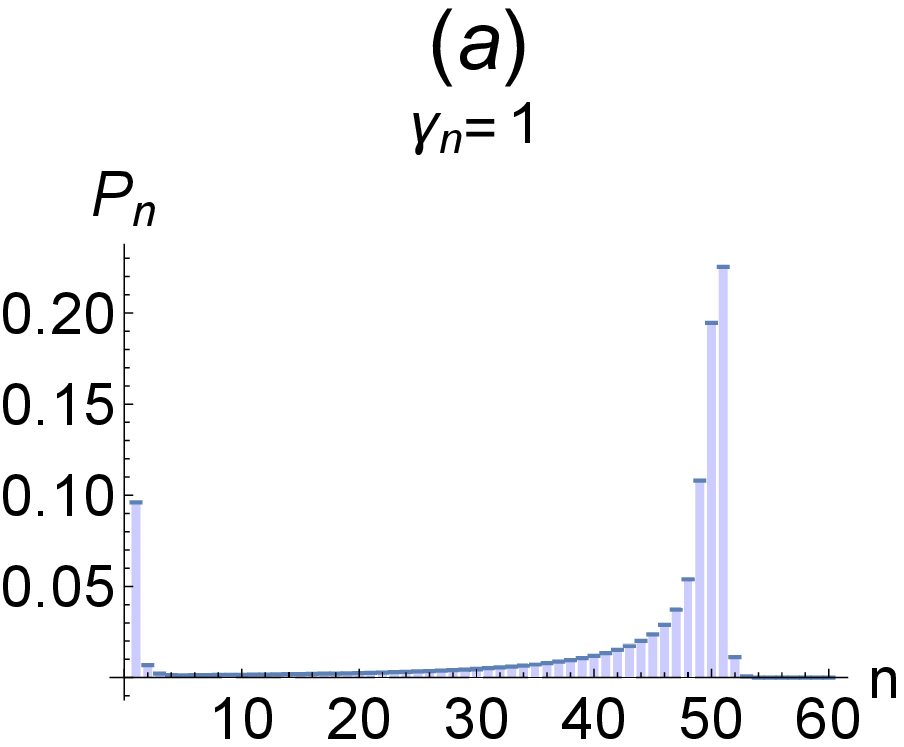}
\includegraphics[width=4.25cm]{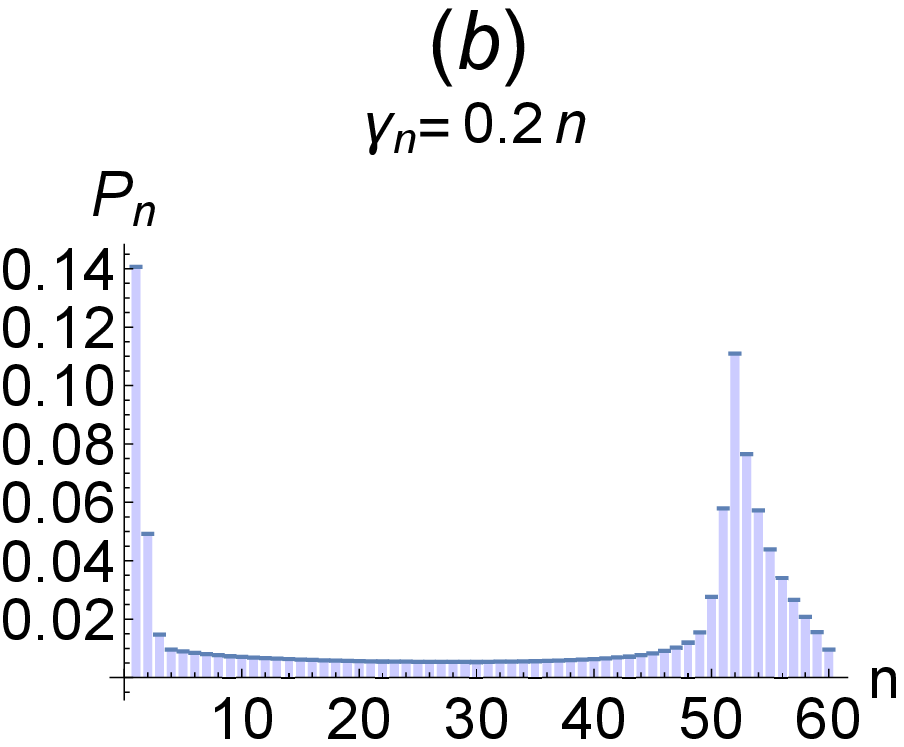}
\includegraphics[width=4.25cm]{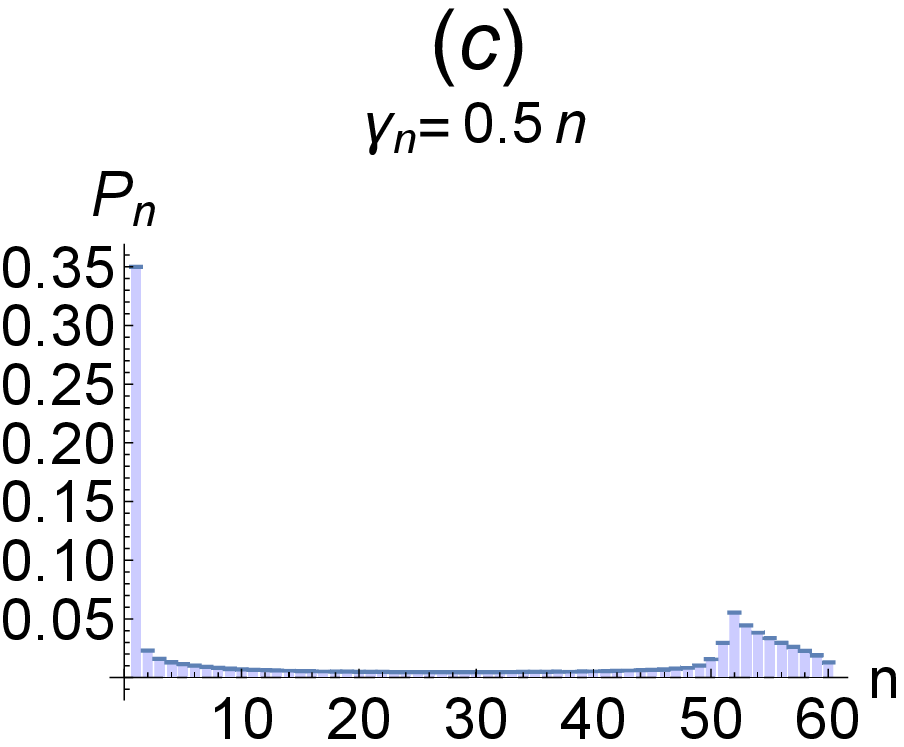}
\includegraphics[width=4.25cm]{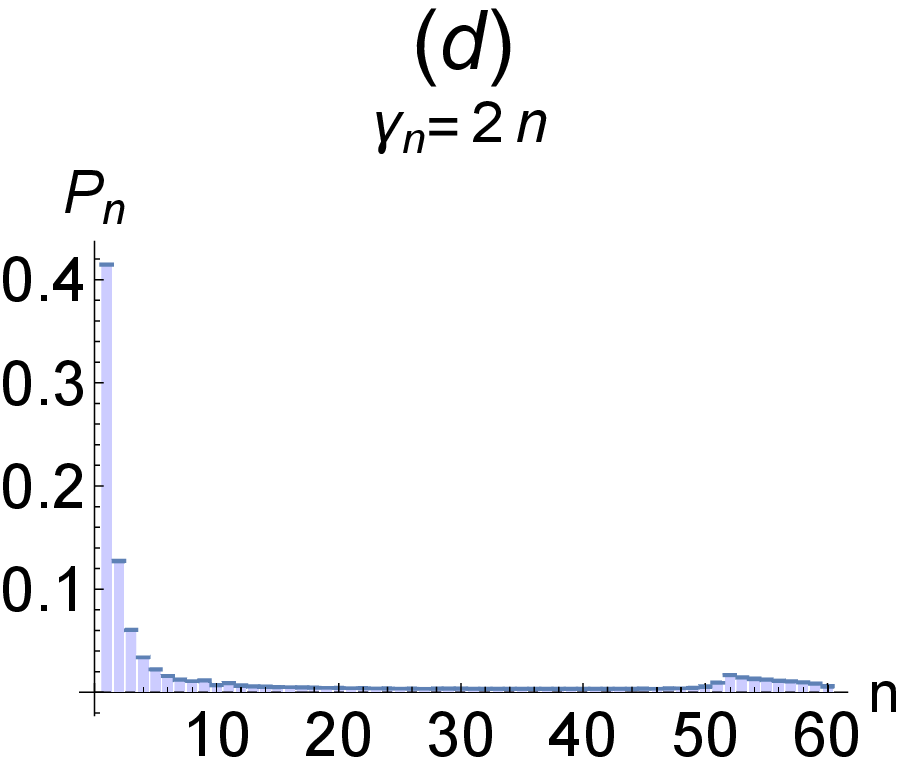}
\caption{The distribution of the decay probabilities for the uniform (a) and nonuniform (b,c,d) loss rates. The ratio$\ds{\frac{P_1}{P_{min} }}$ is bigger than one for all figures ($18$, $25$, $76$ and $124$ for (a,b,c) and (d), respectively). However, the ratio $\ds{\frac{P_1}{P_S}}$ is smaller than $1$ only for (a). The edge burst is enhanced with increasing $\gamma$: 14, 35 and 40 percentages of the total decay occur at the left edge for (b), (c) and (d), respectively. The left-right asymmetry in (b,e) cannot be attributed to the NHSE since they have no skin localizations at the left edge. The decay probabilities don't fall sharply in the right of the starting point in (b-d) as opposed to the case in (a). The parameters are $t_1=0.3$, $t_2=0.5$, $S=50$ and $N=60$. }
\end{figure}
In each plot in Fig.2, the distribution of $\ds{P_n}$ is left-right asymmetric. Besides, in the right of the starting point ($n>S$), $\ds{P_n}$ falls sharply to be almost zero value in (a) because of the strong NHSE, whereas it falls softly in (b-d). In fact, an extensive number of eigenstate are no longer localized at the left edge when the loss rates are nonuniform (\ref{iyrl0dsf}) and hence we can not attribute the left-right asymmetry in Fig. 2 (b-d) directly to the NHSE. This is in stark contrast to the case considered in \cite{in2}, in which the authors conclude that the NHSE is necessary for the edge burst. Below, we explore this issue in more detail.\\
Let us find the energy spectra for the ring and open configurations. The ring configuration is the one for which the left and right edges of the lattice are connected ($\ds{ \psi^{A}_{N+1} =\psi_1^{A} }$, $\ds{  \psi^{A}_{0} = \psi^{A}_{N} }$, $\ds{ \psi^B_{N+1} =\psi_1^B}$, $\ds{  \psi^{B}_{0} = \psi^{B}_{N} }$) and the open configuration is the one with two open edges ($\ds{ \psi^A_{0} =\psi_{N+1}^A=\psi^B_{0} =\psi_{N+1}^B=0 }$). The ring lattice has extra couplings between the edges than the open lattice, but this small change leads to a drastic spectral change in the case of the uniform loss rate $\gamma_n=\gamma$, which is an indication of the NHSE \cite{PRX}.  The spectrum for the ring lattice makes a loop in the complex energy plane, whereas the spectrum for the open lattice is placed inside this loop (Fig.3 (a)). Similarly, the corresponding eigenstates have different characters. An extensive number of eigenstates are localized at one edge of the open lattice, whereas the eigenstates are extended for the ring lattice. However, such a drastic spectral difference doesn't arise in the case of the nonuniform loss rate (\ref{iyrl0dsf}). The spectra for the ring and open lattices are T-like shaped (Fig.3 (b)). The eigenvalues for the ring lattice makes a tight loop along the $E_R$-axis (inset of Fig.3 (b)), while almost coincide with the eigenvalues for open lattice along the $E_I$ axis. This loop is compressed along the vertical direction as $N$ and $\gamma$ increase and becomes a line along $E_R$-axis for the semi-infinite boundary condition ($N\rightarrow\infty$), meaning that the spectrum perfectly becomes T-shaped. We can also study the NHSE by exploring the localization character of the eigenstates for the open lattice. In order to quantify the skin localization, we use the averaged mean displacement over all energy eigenvalues for both sublattices 
\begin{figure}[t]
\includegraphics[width=4.20cm]{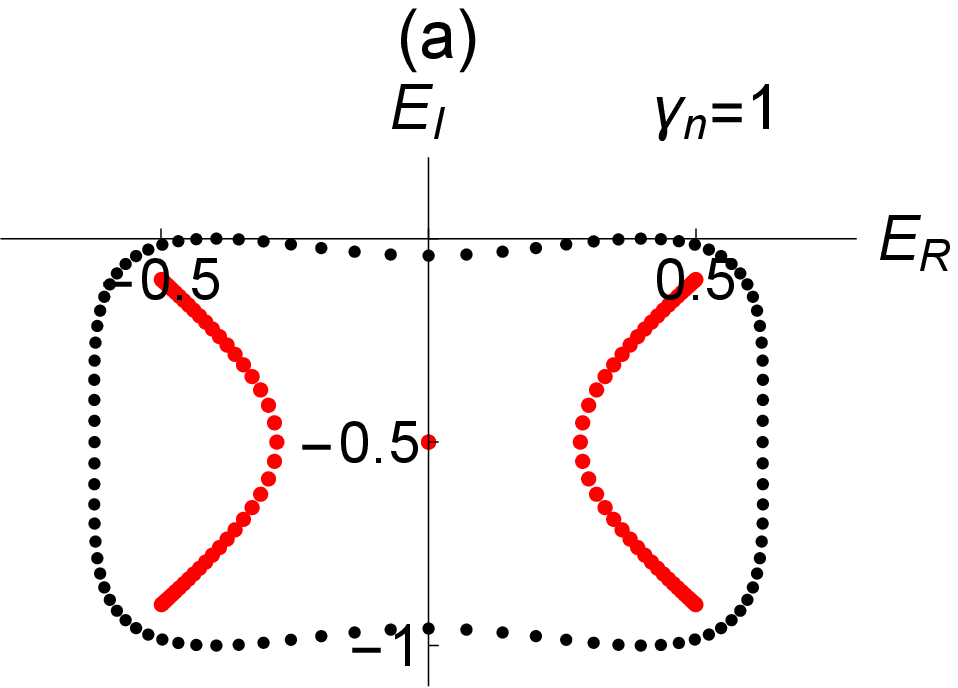}
\includegraphics[width=4.30cm]{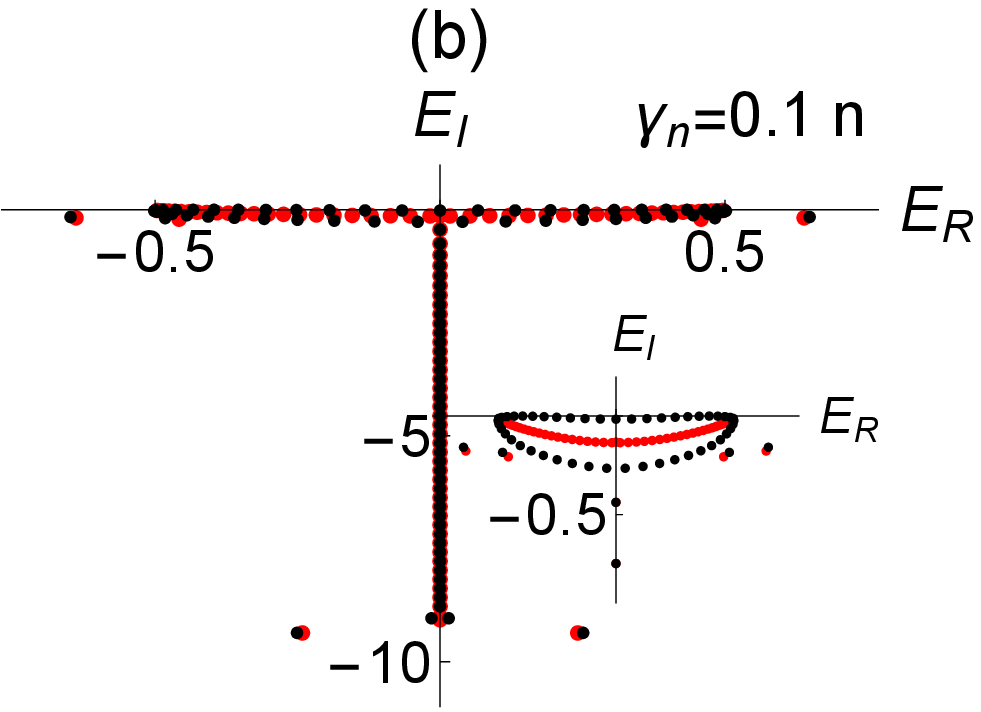}
\caption{The energy spectra for the uniform (a) and nonuniform loss rates (b), where the black and red points are for the ring and open geometries, respectively. The spectra for open and ring lattices are drastically different only for the uniform lattice, indicating that NHSE occurs  for the uniform one. The spectra are $T$-like shaped for the nonuniform lattice, and close the imaginary gap. In the inset, we show the spectra close to the $E_R$-axis. The parameters are $t_1=0.3$, $t_2=0.5$ and $N=96$.  }
\end{figure}
\begin{figure}[t]
\includegraphics[width=4.25cm]{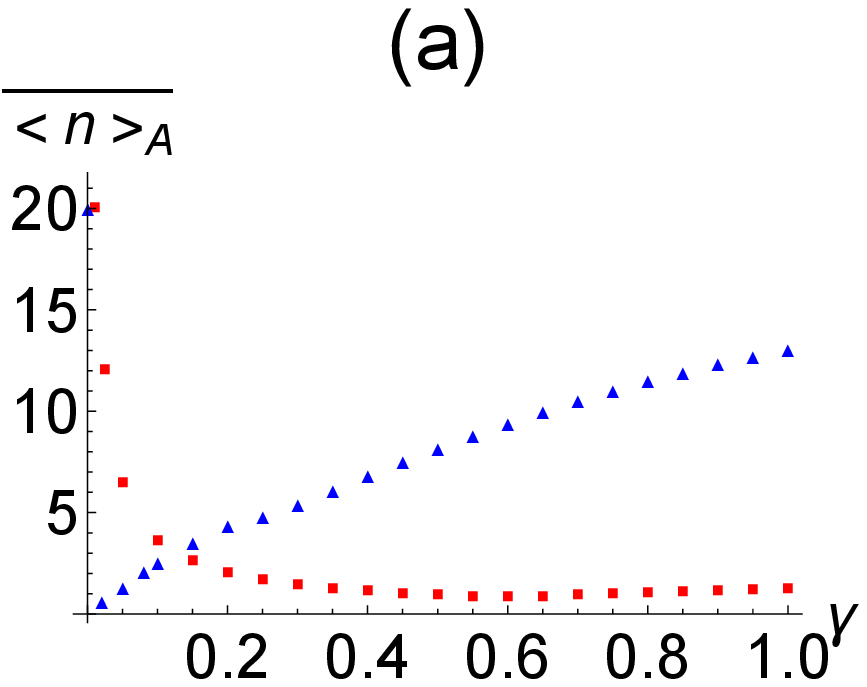}
\includegraphics[width=4.25cm]{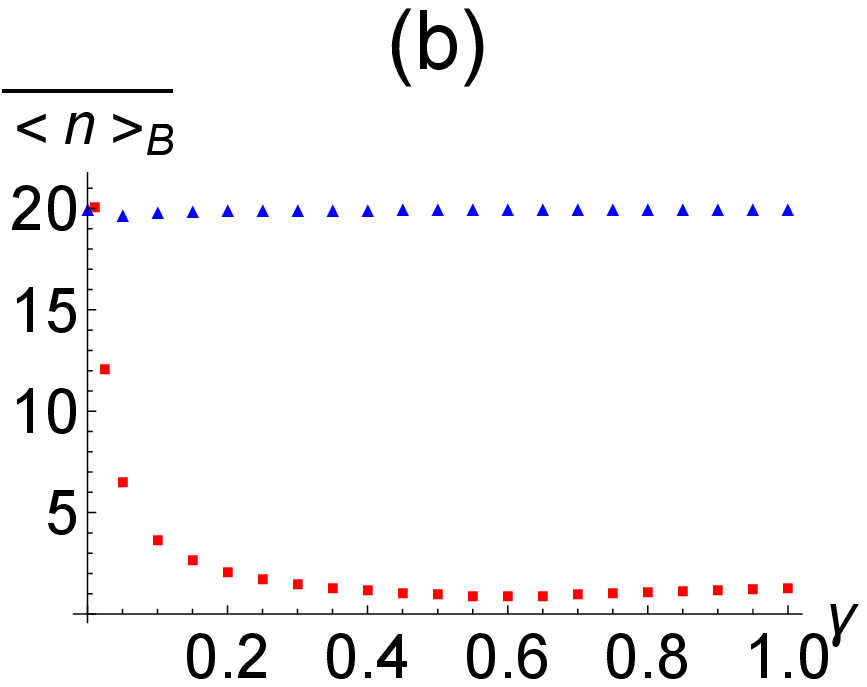}
\caption{The averaged mean displacements $ \overline{<n>_A } $ and $ \overline{<n>_B } $ as a function of $\gamma$ for the $\mathcal{A}$ and $\mathcal{B}$ sublattices, respectively. They decrease for the open lattice with the uniform loss rates, $\gamma_n=\gamma$ (in red) and become very small unless the loss rate is small, indicating that NHSE occurs. However, an extensive number of eigenstate localization doesn't occur at the left edge for the nonuniform loss rates, $\gamma_n={\gamma}n$ (in blue) since the averaged mean displacements are not small for both sublattices. The parameters are $t_1=0.3$ and $t_2=0.5$, $N=40$.  }
\end{figure}
\begin{eqnarray}\label{tcniHGlk2} 
 \overline{<n>}_{A,B}   =\frac{1}{N}\sum_{E}   \sum_n   ~ n~|\phi_n^{  A,B  }|^2 
\end{eqnarray} 
where $\ds{  ~ \phi_n^{A,B} }$ are the normalized stationary solutions: $\ds{ \psi_n^A(t) =e^{-iE t} ~ \phi_n^A }$ and $\ds{ \psi_n^B(t) =e^{-iE t} ~ \phi_n^B }$ with $\ds{  \sum_n   ~|\phi_n^{  A  }|^2 +|\phi_n^{  B  }|^2=1  }$.  A skin state localized at the left edge makes little contribution in this summation, whereas an extended state or a state localized away from the left edge make a high contribution. Therefore, the averaged mean displacement is much smaller than $\ds{N}$ if the system is in the skin phase at which an extensive number of eigenstates are tightly localized at the left edge. We plot the averaged mean displacements as a function of $\ds{\gamma }$ for the uniform (in red) and nonuniform (in blue) loss rates in Fig. 4. In the case of the uniform loss rate, it decreases with $\ds{\gamma}$ in both sublattices since the states get localized more tightly at the left edge due to the NHSE. However, this is not the case for the nonuniform loss rate, and moreover both sublattices show different behaviors. We mote that an extensive number of eigenstates can be tightly localized at the left edge only in the $\mathcal{A}$ sublattice if $\gamma$ is small. We numerically see that eigenstates in the $\mathcal{A}$ sublattice, $\ds{\phi_n^A}$, become extended for large values of $\gamma$. The system as a whole is not in the skin phase since $ \ds{   \overline{<n>}_{B}   } $ is nearly equal to $N/2$ at $\gamma=0$ and this changes slightly with $\ds{  \gamma }$. To be more precise, the eigenstates in the $\mathcal{B}$ sublattice form a complex Wannier-Stark ladder, i. e., the system gives rise to an almost equidistantly spaced energy spectrum in the imaginary axis (the corresponding potential can be thought of as complex electric potential, leading to Wannier-Stark localization \cite{wanni}). Therefore, localization occurs around each lossy lattice points, leading an insulating behavior in $\mathcal{B}$ sublattice in the context of transport. Therefore the quantum walker is mostly transported in $\mathcal{A}$ sublattice (conducting) to the left edge from which it decays. The degree of localization (delocalization) in $\mathcal{B}$ ($\mathcal{A}$) sublattice increases with $\gamma$, so the edge burst is enhanced with increasing $\gamma$, and the ratio $P_1/P_S$ increases with $\ds{\gamma}$ as we numerically see in Fig. 2 (b-d). To this end, we think that the edge burst can be seen as long as $\mathcal{B}$ sublattice has an insulating character. As an example, we consider random values of $\gamma_n$ at which Anderson localization appears in $\mathcal{B}$ sublattice,  and see the edge burst can occur.\\
We next discuss the reason why the left-right asymmetry of $\ds{P_n}$ appears even in the absence of NHSE. We begin to note that the $\mathcal{A}$ and $\mathcal{B}$ sublattices favor opposite propagations even in the Hermitian case ($\ds{\gamma=0}$), implying that the asymmetric behavior of $P_n$ has nothing to do with the non-Hermiticity directly. This is because of the fact that 
the sign of the forward and backward couplings ($it_2/2$ and $-it_2/2$) in $\mathcal{A}$ sublattice is reversed in $\mathcal{B}$ sublattice. This phase difference generates counterclockwise motions in the unit cell such that the  $\mathcal{A}$ and $\mathcal{B}$ sublattices favor motions to the left and right, respectively. In the Hermitian lattice, the quantum walker moves to the left (right) in $\mathcal{A}$ ($\mathcal{B}$) sublattice  in such a way that no net motion is generated. Fortunately, the motion towards right in $\mathcal{B}$ sublattice is suppressed if we introduce losses in $\ds{B}$ sublattice. This leads to asymmetric behavior of the quantum walker in our system, meaning that NHSE is not necessary for the edge burst. Note that this asymmetric behavior disappears if the loop in the unit cell is broken at $\ds{t_1=0}$.\\
\begin{figure}[t]
\includegraphics[width=4.25cm]{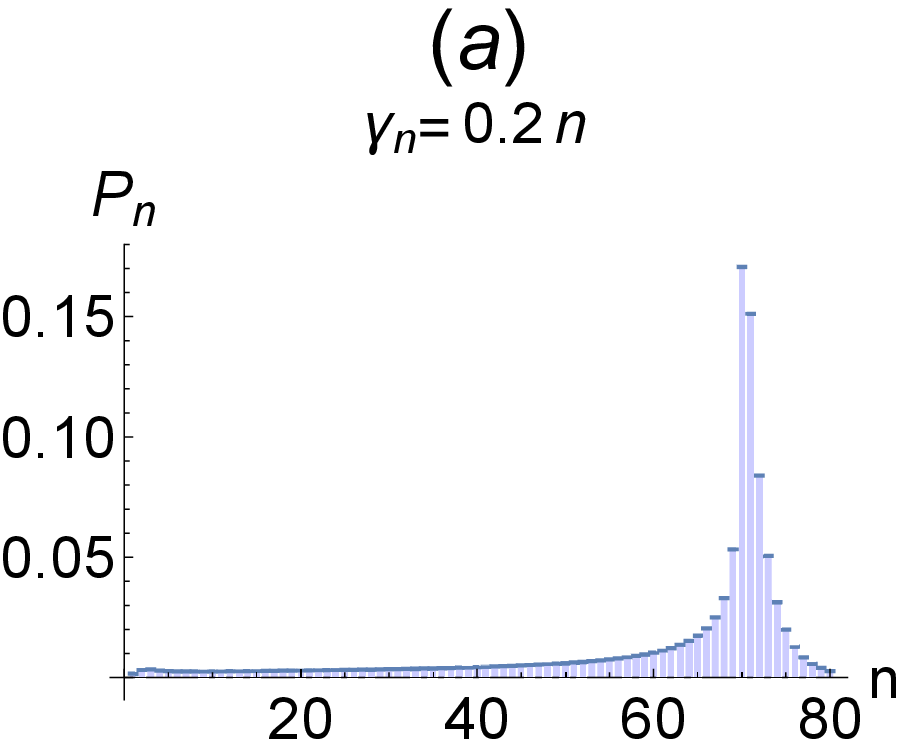}
\includegraphics[width=4.25cm]{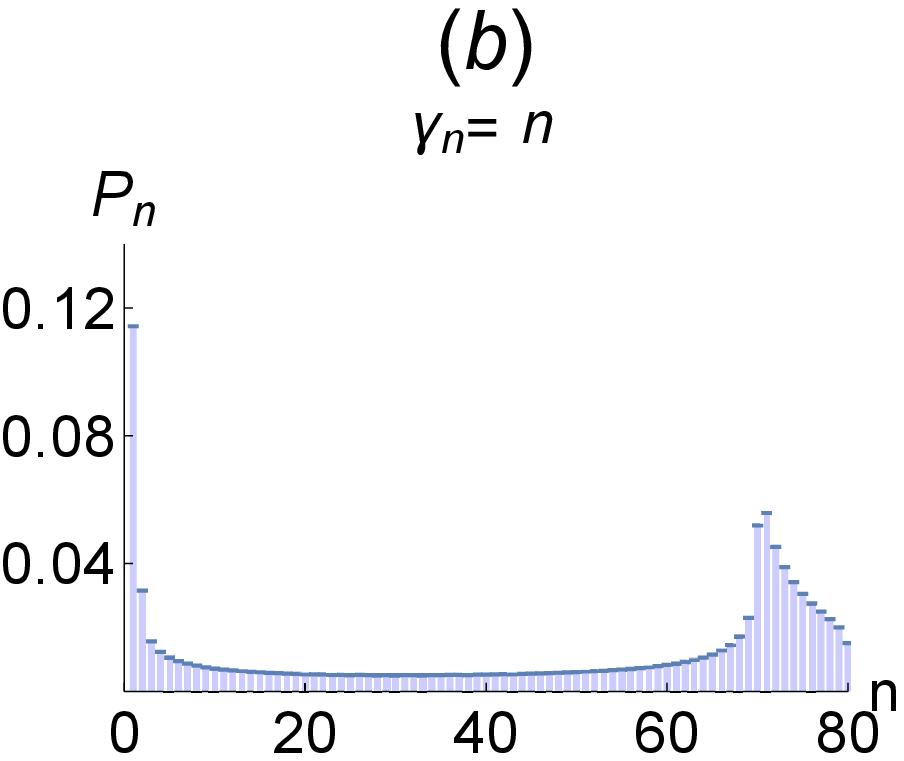}
\includegraphics[width=4.256cm]{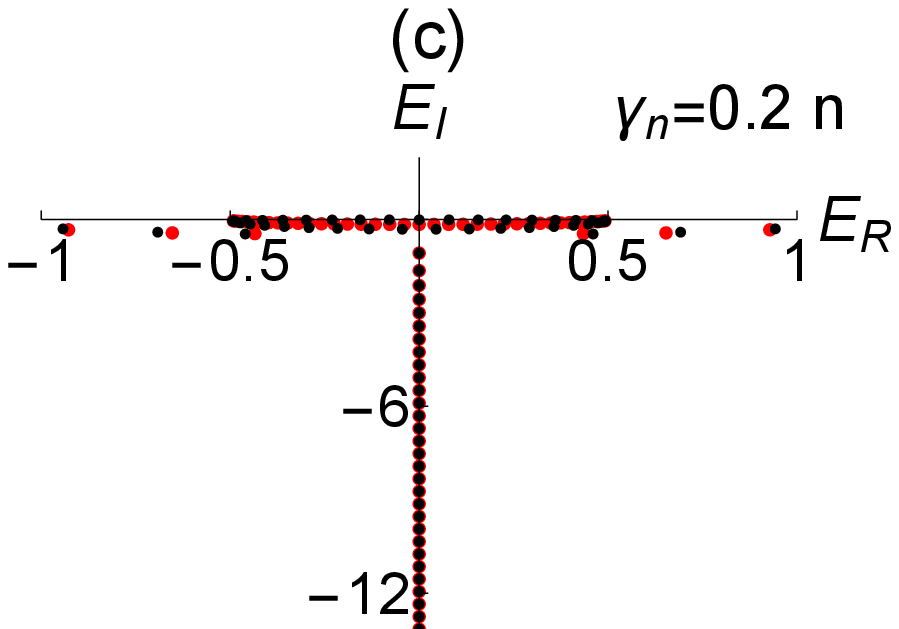}
\includegraphics[width=4.25cm]{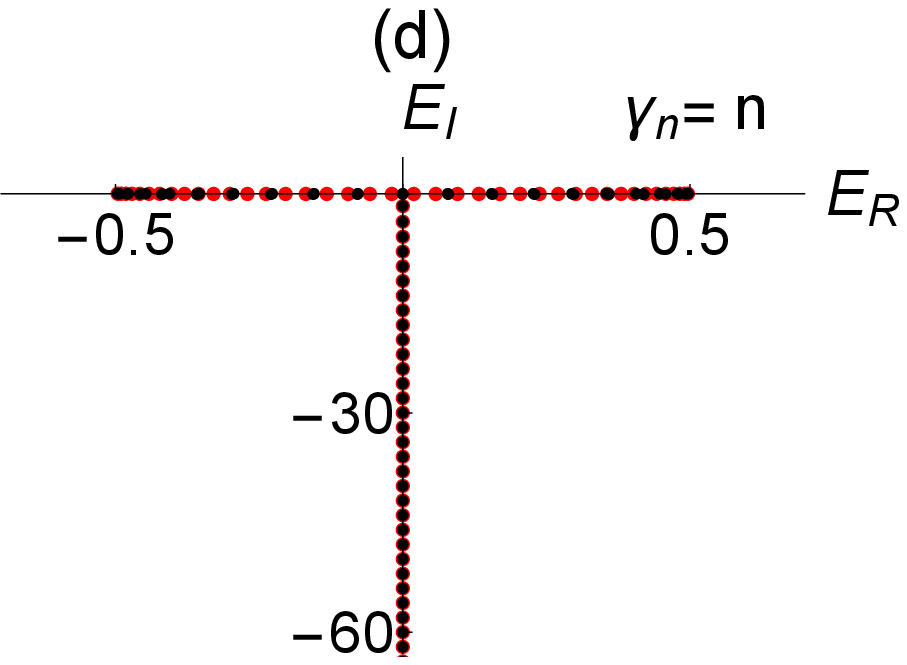}
\caption{$\ds{P_n}$ for two different $\gamma_n$ with $\ds{P_1/P_{min}  =1 }  $ (a) and $\ds{P_1/P_{min}  =23 }  $ (b). The corresponding spectra in the complex plane have $T$-shape structure (c,d). The ring (in black) and open (in red) lattices have almost the same spectra. Both systems practically satisfy the imaginary gaplessness condition (The maximum eigenvalues on the imaginary axis for the ring lattices are $0.009$ and $0.002$ for (c) and (d), respectively. These very small but non-zero values are due to the finiteness of the lattice). The parameters are $t_1=0.7$ and $t_2=0.5$, $N=80$ and $S=70$. }
\end{figure}
As mentioned above, the edge burst for the uniform lattice is thought to be originated  jointly from NHSE and the imaginary gaplessness (the spectrum under periodic boundary conditions touches the real axis and closes the imaginary gap) \cite{in2}. The $T$-shaped spectra in the complex plane seem to satisfy the imaginary gaplessnes condition for the appearance of the edge bursts in our specific examples. However close inspections reveal that there may be some other examples with $T$-shaped spectra but without the edge burst effect. Let us choose $t_1>t_2$. We show two such examples in Fig. 5 with $T$-shaped spectra. The edge burst doesn't appear for a small value of $\gamma$ (a). However, it appears when $\gamma$ is large (b) .\\
A relatively huge peak at the edge in the distribution of the decay probabilities in the bipartite lossy lattices is the evidence of the so-called edge burst. In this Letter, we consider such lattices with the nonuniform loss rate and show that edge burst occur even if an extensive number of eigenstates are not localized at one edge of the system due to NHSE. We discuss that the left-right asymmetry of the decay probabilities can be due to the phase difference of the couplings in the unit cell. We also show that the edge burst may not appear even if the spectrum closes imaginary gap in the complex energy plane. Our paper shows interesting dynamics and can stimulate other researchers to study dynamics of non-Hermitian systems in order to find the real source of the edge burst. \\
H. R. acknowledges the support by the Army Research Office Grant No. W911NF-20-1-0276, NSF Grant No. PHY-2012172 and OMA-2231387.

\end{document}